\documentclass[conference]{IEEEtran}
\IEEEoverridecommandlockouts
\usepackage{cite}
\usepackage{amsmath,amssymb,amsfonts}
\usepackage{algorithmic}
\usepackage{graphicx}
\usepackage{textcomp}
\usepackage{xcolor}
\usepackage{hyperref}
\def\BibTeX{{\rm B\kern-.05em{\sc i\kern-.025em b}\kern-.08em
    T\kern-.1667em\lower.7ex\hbox{E}\kern-.125emX}}
\begin{document}

\title{Dynamic infrared thermography and computed tomography angiography fusion for DIEP flap perforator selection: a novel measurement system
\thanks{This research was funded by the Research Foundation Flanders (FWO) under a Strategic Basic Research (SB) fellowship [1SB9825N] (Clarys Warre), and by the FWO under a grant for Applied Biomedical Research with a Primary Social Goal [T001723N].}
}

\author{
\IEEEauthorblockN{
Warre Clarys\IEEEauthorrefmark{1},
Roanne Van Steen\IEEEauthorrefmark{2},
Simon Verspeek\IEEEauthorrefmark{1},
Filip Thiessen\IEEEauthorrefmark{2},
Gunther Steenackers\IEEEauthorrefmark{1},
}
\IEEEauthorblockA{\IEEEauthorrefmark{1}
InViLab Research Group, University of Antwerp. Antwerp, Belgium\\
warre.clarys@uantwerpen.be; 0000-0002-4109-8839; 0000-0001-9944-520X}
\IEEEauthorblockA{\IEEEauthorrefmark{2}
Department of Plastic Surgery, University of Antwerp. Antwerp, Belgium\\
0009-0005-1935-4620; 0000-0002-4958-3150}
}

\maketitle

\begin{abstract}
Accurate perforator selection is critical for successful deep inferior epigastric artery perforator (DIEP) flap breast reconstruction. While computed tomography angiography (CTA) provides high-resolution anatomical detail, it lacks functional perfusion assessment. Dynamic infrared thermography (DIRT), in contrast, captures superficial rewarming dynamics but lacks objective anatomical validation. This study presents a standardized multimodal measurement system enabling automated fusion of DIRT with CTA-derived perforator coordinates. The framework integrates controlled thermographic acquisition, real-time hotspot detection, metric spatial calibration, and global one-to-one assignment of thermal hotspots to CTA perforators within a predefined periumbilical region of interest. The workflow was evaluated in 12 patients. Of 46 CTA perforators within the region of interest, 84.8\% were successfully linked to thermographic hotspots, with a median spatial offset of 14.74 mm. Automated assignment excluded 76.1\% of detected hotspots due to lack of anatomical correspondence, thereby reducing subjective interpretation. The proposed system enables quantitative multimodal validation of functional and anatomical imaging and provides objective support for perforator selection in clinical practice.

\end{abstract}


\section{Introduction}
Breast cancer remains the most prevalent malignancy worldwide, with a continued rise in incidence projected over the coming decades \cite{WHO2022}. In parallel, breast reconstruction rates after mastectomy have increased substantially, and autologous techniques are increasingly preferred due to superior long-term aesthetic and functional outcomes \cite{ASPS2022, Myers2021}. Among these, the deep inferior epigastric artery perforator (DIEP) flap is widely regarded as the gold standard for abdomen-based breast reconstruction.

Successful DIEP flap reconstructions depend on accurate preoperative identification and selection of perforating vessels, as suboptimal perforator choice increases operative time, flap complications, and postoperative morbidity \cite{Thiessen2019, Wade2018}. Consequently, robust and reproducible preoperative perforator mapping is a key element in modern reconstructive planning.

Computed tomography angiography (CTA) is widely considered the clinical reference standard for perforator mapping, offering high-resolution visualization of perforator location, course, and caliber. Meta-analytic evidence suggests that CTA-guided planning improves localization and surgical outcomes, albeit with variability in protocols and radiation exposure concerns \cite{Rupra2025_CTAmeta, Kim2025_CTAplanning}. Alternative modalities such as color Doppler ultrasound have been comparatively evaluated, highlighting trade-offs between anatomical precision, operator dependence, and clinical workflow integration \cite{Bajus2023_CDUvsCTA}.

Dynamic infrared thermography (DIRT) leverages superficial rewarming patterns after controlled cooling to assess perfusion characteristics non-invasively and without ionizing radiation \cite{Moderhak2016}. In the context of DIEP planning, thermography has been investigated for preoperative perforator detection using both research-grade and portable devices \cite{Alberti2025_ThermoDevice}, and as part of multi-modal perfusion studies alongside indocyanine green fluorescence and hyperspectral imaging \cite{Kleiss2024_PerfusionMulti}. However, thermographic interpretations are typically qualitative and rely on visual assessment of hotspots without quantitative anatomical confirmation.

Despite these advances, the literature currently lacks a standardized framework that quantitatively integrates functional thermographic data with anatomical CTA reference information. Approaches using projected augmented reality or extended reality have been proposed to visualize complementary information on the patient, but they generally do not implement automated spatial co-registration or objective multimodal validation \cite{Meier2024, Lacey2025_XRreview}. Moreover, published thermography studies rarely provide algorithm-driven matching against external ground truth, instead depending on subjective interpretation \cite{Clarys2026_B}.

In this work, we present a measurement-driven multimodal fusion system that enables automated linking of DIRT thermal hotspots with CTA-derived perforator coordinates. Our framework combines controlled thermographic acquisition, real-time computational processing, and global one-to-one assignment to objectively associate functional perfusion signals with anatomical targets within a clinically defined periumbilical region of interest, with evaluation in a clinical patient cohort.

\section{Materials and Methods}

\subsection{Study Design and Population}

This study assessed a multimodal imaging workflow combining DIRT with preoperative CTA for perforator mapping in DIEP flap planning. Consecutive adults scheduled for unilateral or bilateral DIEP flap breast reconstruction were eligible. 12 patients were enrolled (mean age 50.7 ± 11.5 years; mean BMI 27.2 ± 3.8 kg/m$^{2}$). Ethical approval was obtained (Belgian registration number B3002024000116), and all participants provided written informed consent.

\subsection{CTA Acquisition}

All patients underwent preoperative abdominal CTA according to institutional protocol. Arterial-phase CTA images were acquired at 0.6–1.0 mm slice thickness with standard multiplanar reconstructions. A dedicated radiologist or experienced reconstructive surgeon identified relevant perforators of the deep inferior epigastric artery and recorded their coordinates relative to the umbilicus in a standardized axial coordinate system (left–right and cranio-caudal offsets in millimeters). These coordinates served as anatomical reference points in the DIRT-CTA fusion workflow.

\subsection{DIRT Measurement Protocol}

Preoperative DIRT was performed with patients in supine position to evaluate perforator activity. System performance and calibration were validated separately; only patient data are reported here.

The dynamic thermography protocol followed established DIEP flap-DIRT procedures using a controlled cold challenge. After camera positioning with the umbilicus centered and in focus, a baseline thermal sequence was acquired. The abdominal skin was then uniformly cooled for 2 min with a standardized method (sterile bag filled with water at 5~\textdegree C). Immediately upon removal, the rewarming phase was recorded continuously for 5 min at 50 Hz, with real-time visualization and hotspot detection provided by the acquisition software.

As the analysis relies on relative temperature differences and rewarming dynamics rather than absolute temperature values, the influence of ambient environmental factors on hotspot detection is limited under stable acquisition conditions.

\subsection{Hardware}

\subsubsection{Multimodal Measurement Head}

DIRT was acquired using a research-grade long-wave infrared camera (FLIR A655sc, Teledyne FLIR, USA). The camera operates in the 7.5–14~\textmu m spectral range and provides radiometric images at a resolution of 640$\times$480 pixels and a frame rate of 50~Hz. The thermal sensitivity (NETD) is below 30~mK at 30~\textdegree C. During clinical measurements, the camera-to-skin working distance was maintained between 0.8 and 1.3~m.

To enable metric scaling and robust geometric alignment, a time-of-flight (ToF) depth camera (ifm O3D303, ifm electronic, Germany) was rigidly coupled to the thermal camera. The ToF sensor provides per-pixel distance measurements over a working range of approximately 0.3–8~m, with a depth accuracy of $\pm$ 4~mm. Depth data were streamed synchronously with the thermal images and used to estimate camera-to-skin distance during acquisition.

\begin{figure}[htbp]
\centering
\includegraphics[width=\columnwidth]{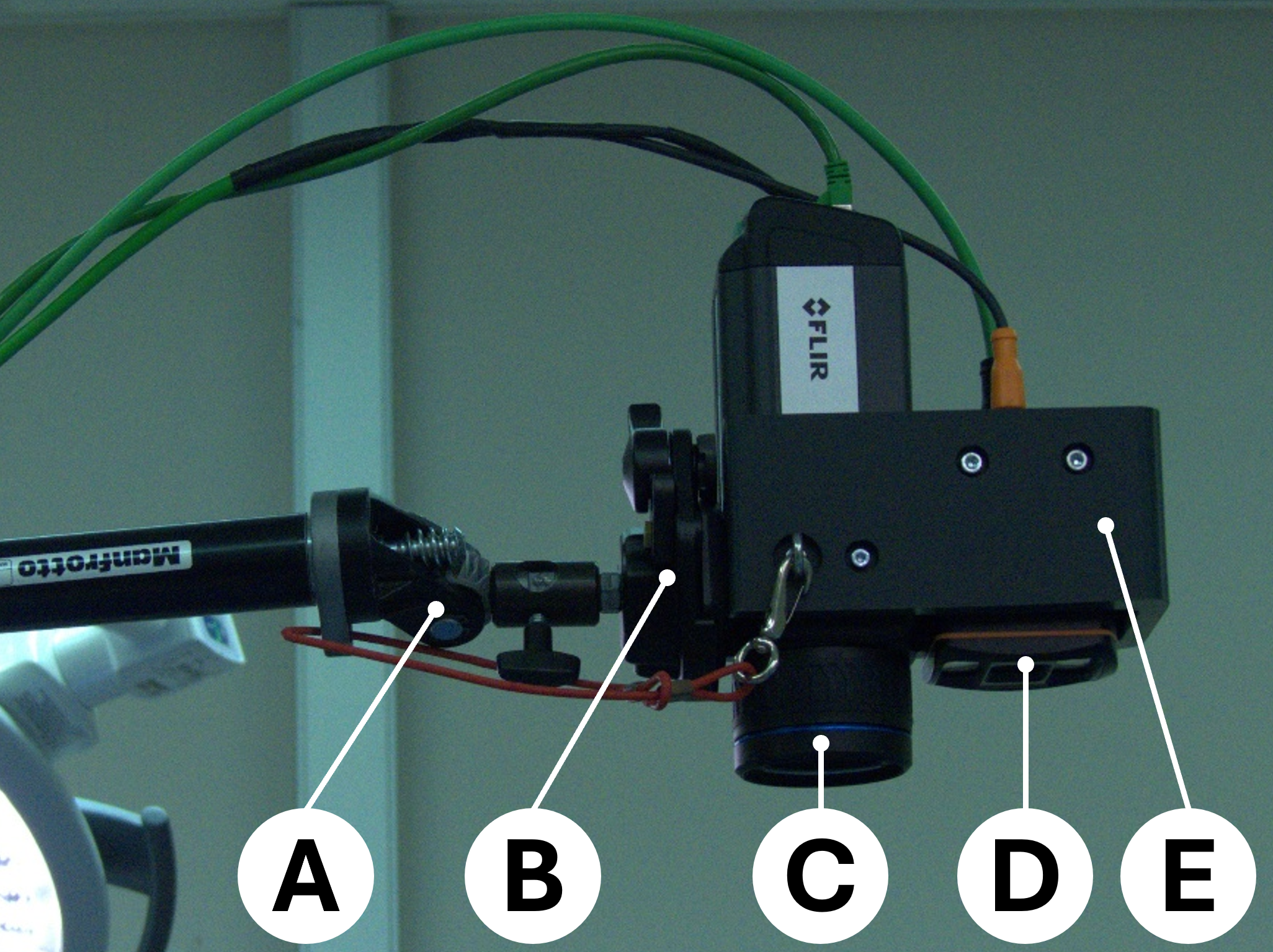}
\caption{Close-up view of the multimodal measurement head: the thermal camera (C) is mounted to the boom stand (A) using a rapid-connect plate (B), while the ToF camera (D) is rigidly coupled to the thermal camera via a custom 3D-printed ASA bracket (E).}
\label{fig:setup_close}
\end{figure}

Both sensors were mechanically integrated into a single measurement head using a custom 3D-printed mounting bracket, ensuring a fixed and reproducible relative pose between the thermal and depth cameras (Fig.~\ref{fig:setup_close}). This rigid coupling is essential for stable cross-modal calibration and consistent spatial mapping across measurements.

\subsubsection{Mechanical Mounting and Clinical Deployment}

The multimodal measurement head was interfaced to the support structure using a Manfrotto~625 rapid--connect plate (Manfrotto, Cassola, Italy). This quick-release interface enabled repeatable attachment while preserving sensor alignment.

\begin{figure}[htbp]
\centering
\includegraphics[width=\columnwidth]{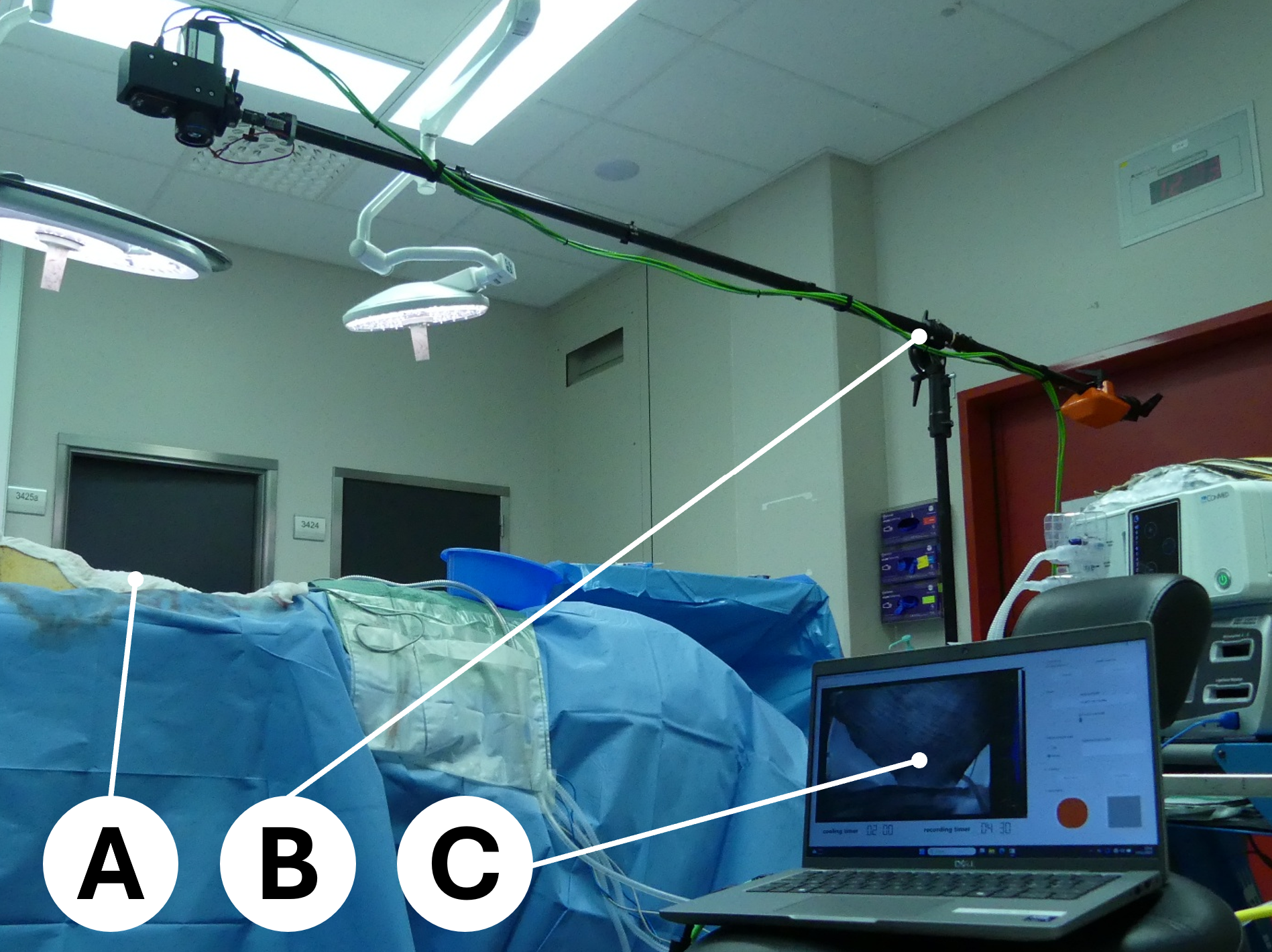}
\caption{Complete measurement setup mounted on the boom stand (B) and positioned above the abdominal region of the patient (A), connected to the acquisition workstation (C).}
\label{fig:setup_total}
\end{figure}

In the clinical setup, the system was typically mounted on a Manfrotto~085BS light boom stand (Manfrotto, Cassola, Italy), allowing stable overhead positioning above the patient in supine position. This configuration ensured a fixed viewing geometry while keeping the floor area around the operating table unobstructed, facilitating integration into routine clinical workflow (Fig.~\ref{fig:setup_total}).

\begin{figure}[htbp]
\centering
\includegraphics[width=\columnwidth]{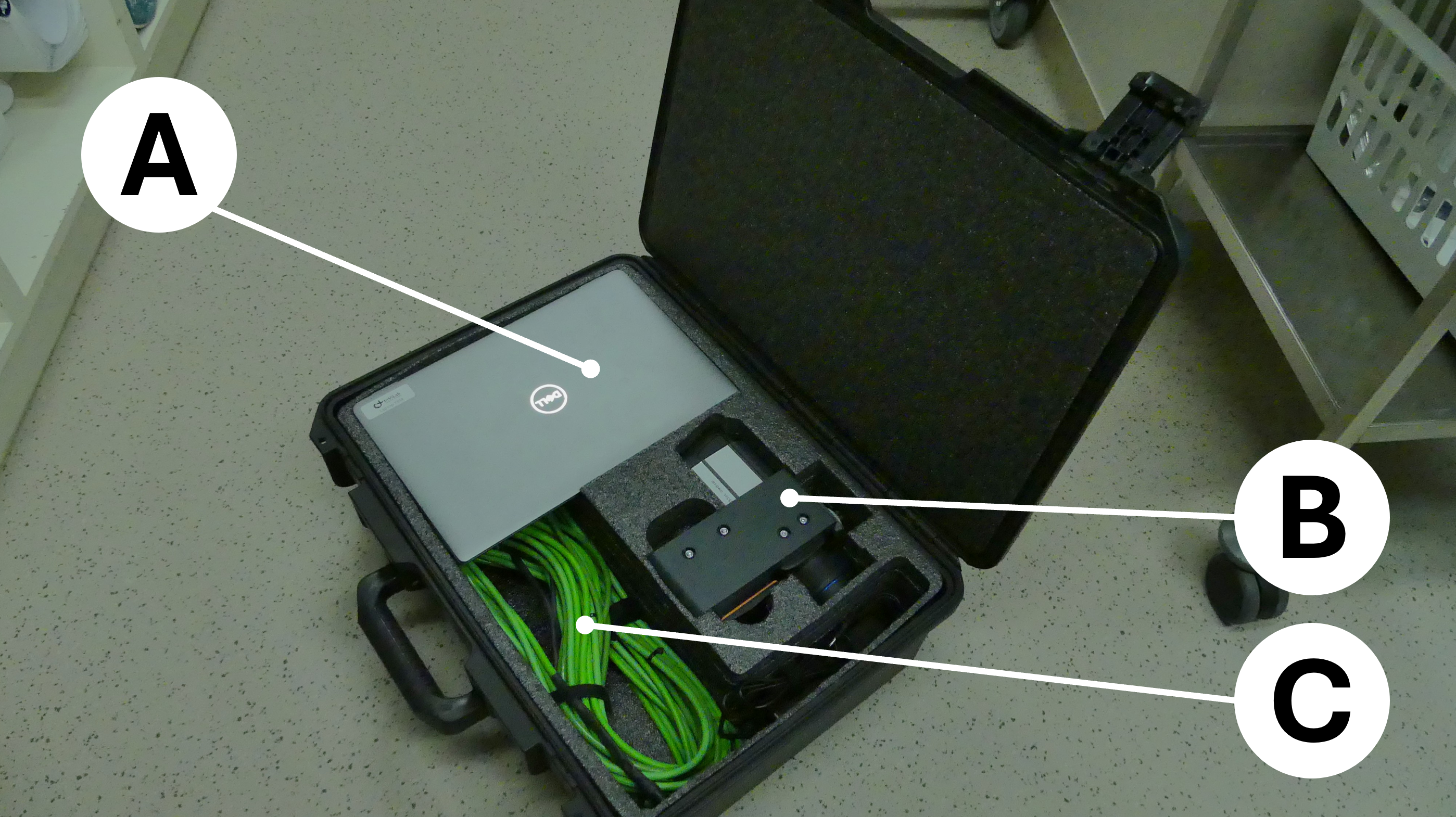}
\caption{Custom flight case with tailored compartments for the measurement head (B), acquisition workstation (A), and cabling (C).}
\label{fig:flight_case}
\end{figure}

For transport and storage, the complete measurement head, mounting components, and cabling were housed in a dedicated flight case with tailored cut-outs. This ensured mechanical protection of the sensors and enabled reproducible system deployment across measurement sessions (Fig.~\ref{fig:flight_case}).

\subsubsection{Acquisition Workstation}

Image acquisition, visualization, and real-time processing were performed on a mobile workstation (Dell Precision 3490, Dell Technologies, USA) running Windows~11 Pro. The workstation was equipped with an Intel Core Ultra processor, dedicated NVIDIA RTX graphics, and 32~GB of RAM, providing sufficient computational capacity for simultaneous thermal streaming, depth processing, hotspot detection, and DIRT–CTA fusion.

\subsection{Software}

\subsubsection{Software Architecture}

\begin{figure*}[t]
\centering
\includegraphics[width=\textwidth]{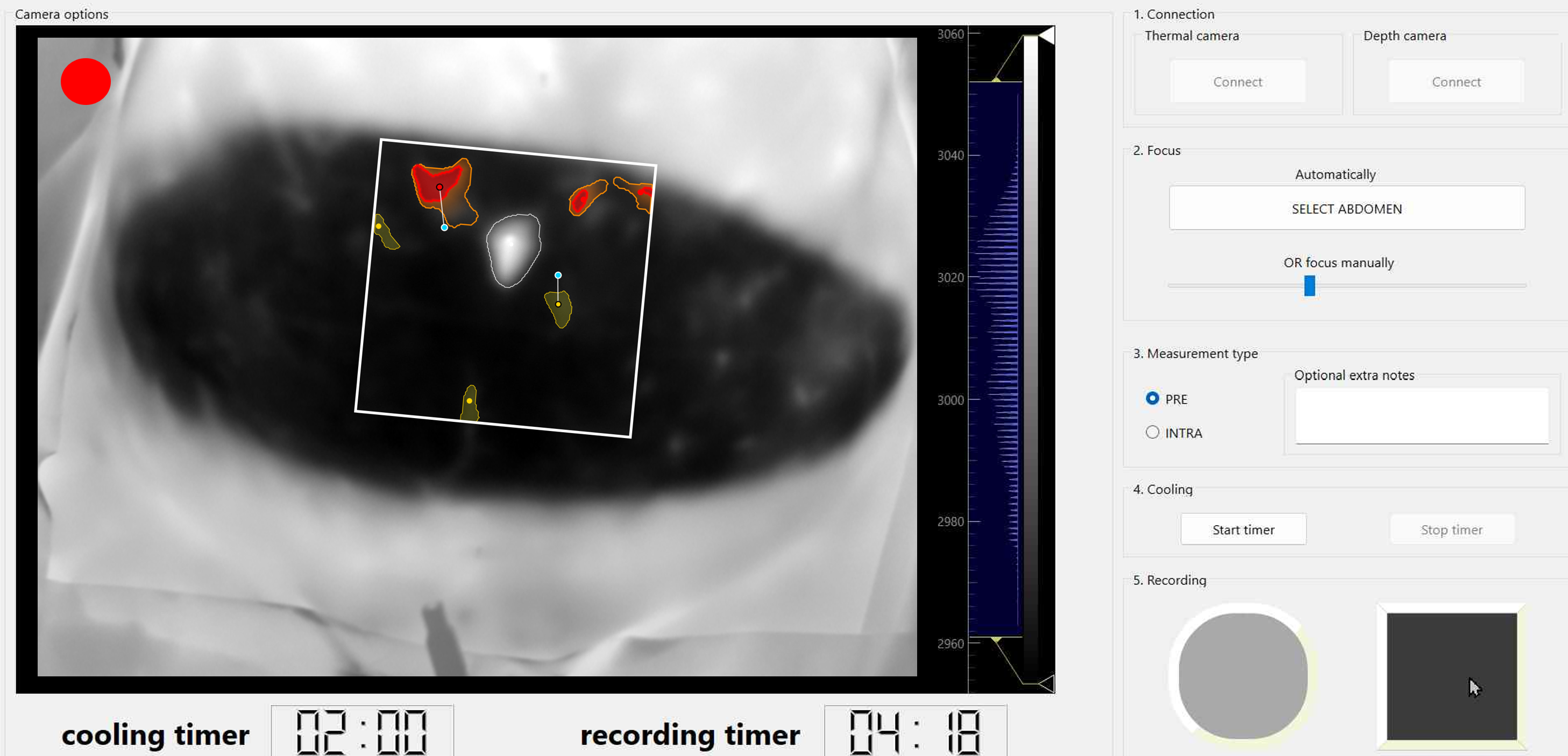}
\caption{Screenshot of the graphical user interface during a measurement. The periumbilical region of interest is automatically defined, detected thermal hotspots are visualized, and CTA-derived perforator locations are overlaid and linked in real time.}
\label{fig:GUI}
\end{figure*}

A custom Python-based acquisition and processing framework was developed to enable real-time analysis of thermal imaging data in combination with CTA-derived perforator information. The software provides a dedicated graphical user interface for pre- and intraoperative use and executes a fully automated processing pipeline (Fig.~\ref{fig:GUI}).

Metric calibration is established before the start of the measurement using a depth-assisted distance estimate. The surgeon selects a point on the abdomen in the live thermal view, which is projected to the depth map using a known cross-modal calibration. The resulting distance is used both to drive thermal camera autofocus and to compute real-time pixel-to-millimeter scaling, ensuring geometric consistency between acquisition and analysis.

Thermal image sequences are streamed into the system and internally normalized to a canonical $(t, y, x)$ representation, which forms the basis for all subsequent spatial processing.

\subsubsection{ROI Definition}

Early in the acquisition, the umbilicus is detected automatically from the initial cooled frames. The abdominal skin in contact with the cooling medium is first identified, after which the umbilicus is segmented as a localized warm structure within this region. A temperature-weighted centroid is computed to define the umbilical origin, which serves as the global anatomical reference for spatial mapping and CTA coordinate transformation.

Perforator analysis is restricted to a fixed periumbilical region of interest (ROI). Anatomical and cadaveric studies consistently report that dominant DIEP perforators cluster within a confined zone surrounding the umbilicus \cite{GagnonBlondeel2006, Chawla2015}. Based on this anatomical evidence and typical clinical harvest zones, the ROI is defined as a square region of 150\,mm × 150\,mm centered on the detected umbilical origin.

Constraining analysis to this predefined square ROI serves both clinical and measurement purposes. Clinically, it focuses detection on the most relevant perforator region. From a measurement perspective, it reduces the spatial search domain, limits boundary-induced thermal artefacts, and improves robustness and stability of the subsequent hotspot-to-perforator assignment. The ROI placement and its relation to detected hotspots and CTA-derived perforators are visualized in the graphical user interface (Fig.~\ref{fig:GUI}).

\subsubsection{Hotspot detection}

Thermal hotspot detection is performed automatically using spatiotemporal analysis of post-cooling rewarming dynamics. Temporal difference images are computed between an early baseline window and later time points in the acquisition, emphasizing localized temperature increases associated with underlying perforators. Within the ROI, adaptive intensity thresholding is applied to the difference images, followed by spatial coherence constraints to retain only connected, physiologically plausible regions. Detected hotspots are classified as primary or secondary based on their temporal behavior. Difference images capturing early rewarming are used to detect primary hotspots, while later difference images reveal secondary, more diffuse rewarming patterns (Fig.~\ref{fig:linking}).

For each hotspot, an intensity-weighted centroid is computed to provide a spatial representation, even in the presence of irregular or extended hotspot morphology. Secondary hotspots that spatially connect to primary regions are treated as extensions, while isolated secondary hotspots are retained as independent candidates.

\subsubsection{DIRT-CTA fusion}

CTA perforator locations, expressed as millimeter offsets from the umbilicus, are mapped into thermal image coordinates using the established pixel-to-millimeter scaling.

Linking between CTA perforators and detected thermal hotspots is formulated as a bipartite one-to-one assignment problem. For each CTA perforator–hotspot pair, the Euclidean distance in physical units is combined with a normalized hotspot brightness measure to compute a composite matching score, with anatomically implausible pairs beyond a 30~mm distance threshold excluded from consideration and a preference given to primary hotspots.

The optimal global assignment is obtained using a linear sum assignment (Hungarian) algorithm, maximizing the total matching score under one-to-one constraints such that each CTA perforator and each hotspot is used at most once. This approach avoids greedy or ambiguous associations and yields spatially and physiologically consistent matches.

\section{Results}

All measurements were successfully completed in the clinical environment without technical failures. The complete measurement setup, including camera mounting, calibration, and software initialization, could be installed and removed within approximately 5 minutes.

\subsection{DIRT–CTA Linking Performance}

\begin{figure}[htbp]
\centering
\includegraphics[width=\columnwidth]{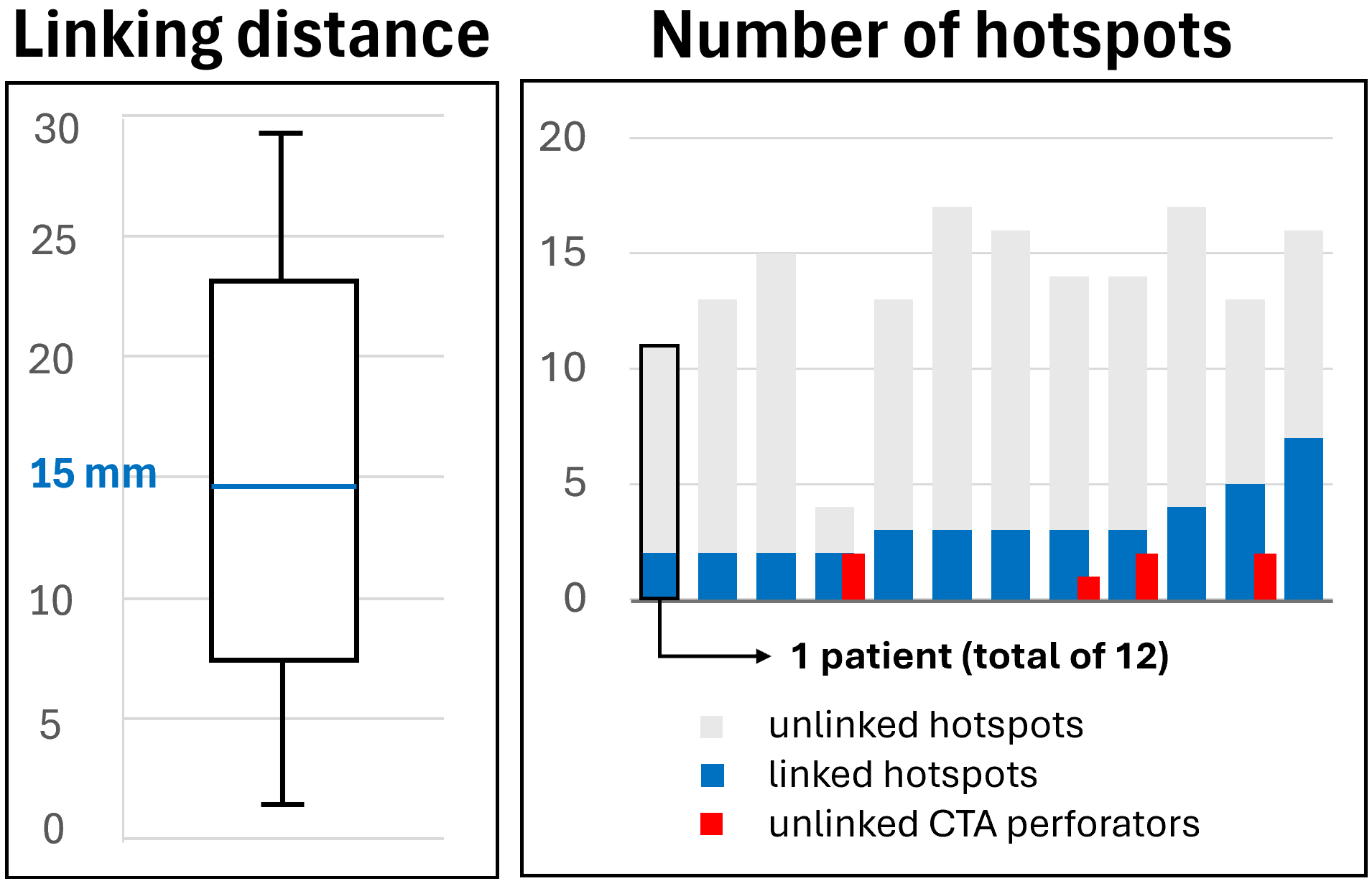}
\caption{Overview of hotspot detection and CTA linking results across all patients. A total of 23.9\% of detected thermal hotspots were linked to CTA perforators, while 76.1\% remained unlinked. These unlinked hotspots included small or peripheral responses, as well as hotspots without a clear anatomical correlate. Overall, 84.8\% of CTA perforators were successfully linked to a thermal hotspot, with a median linking distance of 14.74~mm.}
\label{fig:hotspots}
\end{figure}

Across the 12 included patients, 46 CTA-derived perforators were identified within the predefined periumbilical ROI and were treated as anatomical reference for performance evaluation. Of these, 39 were successfully linked to a thermographic hotspot, corresponding to a sensitivity of 84.8\% (95\% CI: 71.1–92.4\%) (Fig.~\ref{fig:hotspots}).

In 8 of the 12 patients, all CTA perforators within the ROI were successfully linked. In the remaining four patients, linking was partial. In three of these cases, the clinically most relevant perforators, based on size and intramuscular course, were successfully linked, whereas smaller or less favorable vessels were not. In one patient, two clinically relevant perforators could not be linked. On average, 3.83 ± 1.67 CTA perforators and 13.58 ± 3.38 thermal hotspots were identified per patient.

In total, 163 thermal hotspots were detected, of which 39 were anatomically linked to CTA perforators, yielding a precision (positive predictive value) of 23.9\% (95\% CI: 17.9–31.2\%). The remaining hotspots frequently exhibited small spatial extent, peripheral location near the ROI boundary, late or diffuse rewarming behavior, or occurrence at the edge of the cooled region. Several were attributed to incomplete cooling coverage resulting in false-positive thermal responses. However, a subset of unlinked hotspots could not be readily explained by acquisition artefacts or anatomical misalignment and may correspond to superficial vascular structures not visualized as DIEP perforators on CTA.

Spatial measurement accuracy was quantified as the Euclidean distance between each CTA perforator coordinate and the centroid of its assigned hotspot. The median linking distance was 14.74 mm (IQR 15.32 mm). This offset reflects both geometric mapping uncertainty and physiological surface displacement, as thermographic rewarming centroids may not coincide exactly with the anatomical perforator exit point.

\subsection{Clinical Interpretation of Linked Hotspots}

When CTA perforators were accompanied by qualitative radiological scores, all perforators labeled as favorable were associated with clearly visible, early-appearing primary hotspots. In contrast, perforators with unfavorable scores were either linked to small, secondary hotspots or could not be linked to any hotspot at all. In cases where no radiological perforator scoring was available, the proposed system provided clear functional differentiation between perforators based on hotspot prominence and temporal behavior, supporting informed intraoperative decision-making.

\begin{figure}[htbp]
\centering
\includegraphics[width=\columnwidth]{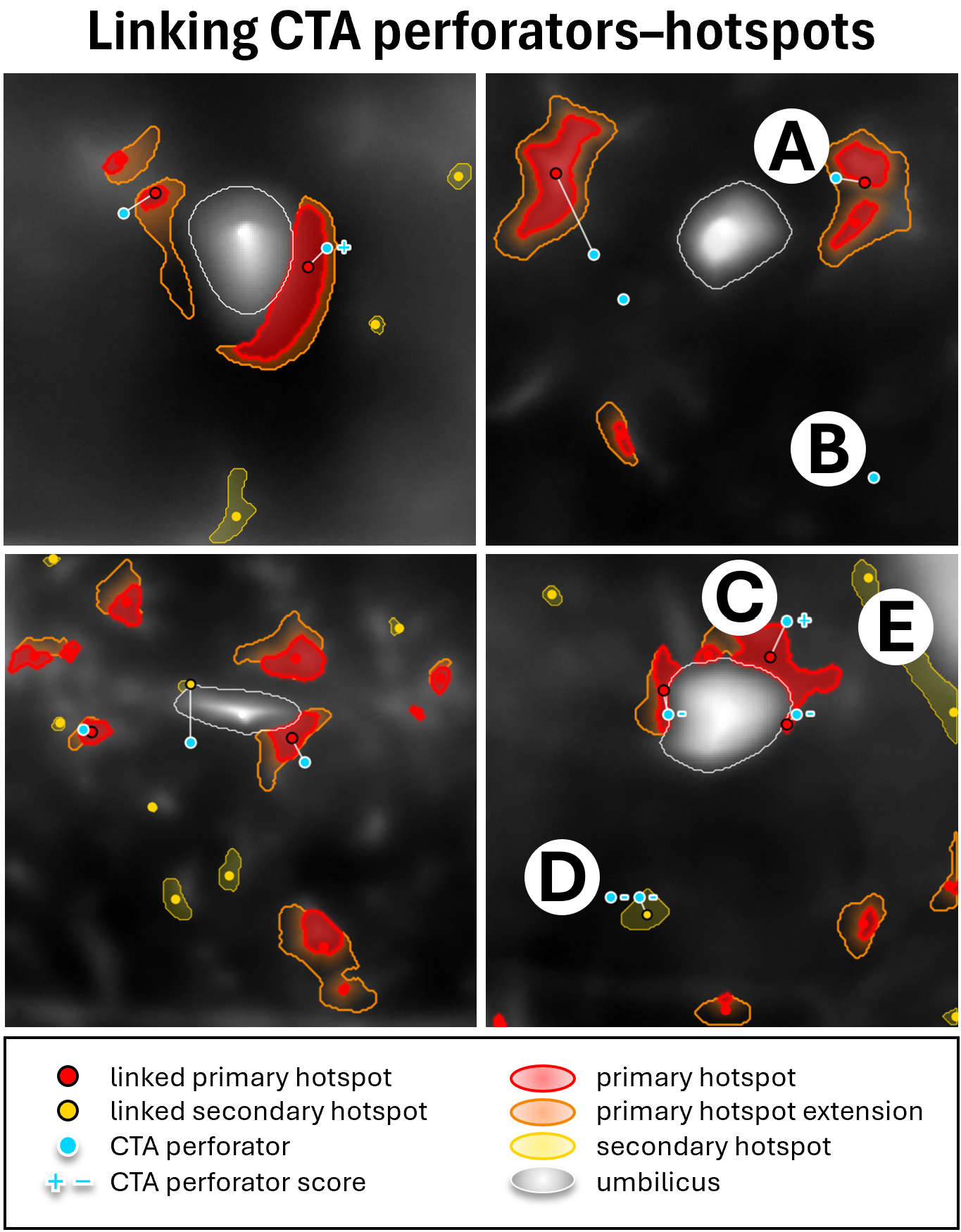}
\caption{Representative examples of automated DIRT–CTA fusion within the periumbilical region of interest (ROI). In cases without radiological perforator scoring (A, B), differences in hotspot prominence and rewarming behavior provide functional information that may support perforator selection. When CTA perforator scores are available (C, D), favorable perforators corresponded to prominent primary hotspots. Incomplete cooling may introduce false-positive hotspots near the ROI boundary (E).}
\label{fig:linking}
\end{figure}

Representative examples of the DIRT–CTA linking results are shown in Fig.~\ref{fig:linking}, illustrating different clinical scenarios and the patient-specific nature of hotspot interpretation.

\section{Discussion}

This study presents a standardized and clinically deployable measurement system that enables automated fusion of DIRT with CTA-derived perforators for DIEP flap planning. The framework combines controlled thermographic acquisition, metric scaling, and global one-to-one hotspot assignment, allowing quantitative evaluation of spatial correspondence between functional rewarming patterns and anatomical perforator coordinates.

A central measurement outcome is the median hotspot-to-perforator linking distance of 14.74~mm. To interpret this value, comparison with contemporary DIRT validation studies is required. Recent clinical investigations correlating DIRT with CTA in DIEP reconstruction have commonly applied spatial matching tolerances in the centimeter range. Meier \textit{et al.} used a predefined 2~cm (20~mm) radius when correlating projected DIRT hotspots with CTA perforators, reporting a 74.5\% DIRT–CTA correspondence within this tolerance \cite{Meier2024}. Similarly, Sonda \textit{et al.} evaluated DIRT–CTA agreement using a 2~cm spatial criterion and reported comparable correlation rates \cite{Sonda2022_Microsurgery}. 

Earlier work by Tenorio \textit{et al.} reported a mean spatial difference of approximately 12~mm between DIRT hotspots and Doppler-identified perforators \cite{Tenorio2011}, while Weum \textit{et al.} described systematic lateral displacement of hotspots relative to CTA-defined perforator locations \cite{Weum2016}. These studies indicate that centimeter-scale spatial variation between functional hotspots and anatomical perforator exit points is intrinsic to DIRT-based mapping.

Importantly, the reported 14.74~mm median offset does not represent pure geometric error, but the spatial separation between an anatomical CTA exit point and a thermographically derived perfusion centroid. Thermal hotspots reflect dynamic superficial rewarming patterns and may be displaced from the perforator trajectory due to oblique subcutaneous courses, perfusion spread, or tissue thickness effects, as described in prior DIRT investigations \cite{DeWeerd2009, Weum2016}.

Beyond spatial distance, 84.8\% of CTA perforators within the predefined ROI were successfully linked to thermographic hotspots. Conversely, 76.1\% of detected hotspots were not anatomically supported by CTA and were excluded through global assignment constraints. Compared to DIRT-only approaches, where hotspot interpretation remains subjective, this multimodal filtering provides structured anatomical validation and reduces ambiguity in perforator selection.

The proposed system augments CTA with quantitative functional information in a reproducible coordinate framework. While CTA visualizes perforator caliber and course, it does not convey dynamic perfusion behavior. The presented approach enables objective association of anatomical and physiological data without replacing CTA as anatomical reference.

Beyond its immediate clinical application, the structured and quantitatively validated DIRT–CTA correspondence also creates a foundation for future artificial intelligence (AI)-based decision support. Standardized multimodal feature extraction and objective hotspot–perforator linking may enable data-driven perforator ranking or relevance prediction in larger patient cohorts.

Several limitations should be acknowledged. The cohort size remains limited, and larger studies are required to further characterize inter-patient variability. The 30~mm anatomical plausibility threshold for assignment was selected in consultation with experienced surgeons; future work will investigate sensitivity of the matching performance to this parameter and further evaluate intraoperative surface validation of hotspot centroids. This study focused on measurement validation; formal usability evaluation was not performed and will be addressed in future work.

Overall, when interpreted in the context of published DIRT–CTA validation studies employing centimeter-scale spatial matching criteria, the reported linking distances and correspondence rates indicate that automated DIRT–CTA fusion can be achieved within clinically plausible spatial variability while providing objective and reproducible multimodal association.

\section{Conclusion}

This study introduced a standardized measurement system for automated fusion of DIRT with CTA-derived perforator information in DIEP flap planning. The proposed framework integrates controlled thermographic acquisition and real-time hotspot detection to objectively associate functional perfusion signals with anatomical perforator ground truth.

Clinical evaluation demonstrated that 84.8\% of CTA-derived perforators within the periumbilical region of interest could be linked to thermographic hotspots, while the majority of thermally active but anatomically unsupported regions were automatically excluded. This multimodal filtering improves interpretability and reduces subjectivity compared to DIRT-only approaches.

By providing surgeons with simultaneous anatomical and functional information in a reproducible workflow, the proposed system supports more informed perforator selection and establishes a practical foundation for objective multimodal validation in reconstructive surgery. Future work will focus on larger clinical cohorts and automated perforator recommendation strategies.

\bibliographystyle{ieeetr}
\bibliography{references}

@misc{WHO2022,
author       = {{International Agency for Research on Cancer}},
title        = {Cancer Tomorrow: Estimated Number of New Cases from 2022 to 2050, Incidence, Both Sexes, Age [0--85+]},
year         = {2024},
url          = {https://gco.iarc.who.int/tomorrow/en/dataviz/tables},
howpublished = {\href{https://gco.iarc.who.int/tomorrow/en/dataviz/tables}{Online}},
note         = {Global Cancer Observatory; GLOBOCAN 2022, version 1.1, 08 February 2024. Accessed: 2024-04-04}
}

@misc{ASPS2022,
author       = {{American Society of Plastic Surgeons}},
title        = {2022 Plastic Surgery Statistics Report},
year         = {2023},
url          = {https://www.plasticsurgery.org/documents/News/Statistics/2022/plastic-surgery-statistics-report-2022.pdf},
howpublished = {\href{https://www.plasticsurgery.org/documents/News/Statistics/2022/plastic-surgery-statistics-report-2022.pdf}{Online}},
note         = {Published: 2023-09-26. Accessed: 2024-04-04}
}

@article{Myers2021,
author       = {Myers, Paige L. and Nelson, Jonas A. and Allen, Robert J., Jr.},
title        = {Alternative flaps in autologous breast reconstruction},
journal      = {Gland Surgery},
year         = {2021},
volume       = {10},
number       = {1},
pages        = {444--459},
doi          = {10.21037/gs.2020.03.16},
pmid         = {33634002},
pmcid        = {PMC7882326}
}

@article{Thiessen2019,
author       = {Thiessen, Filip E. F. and Tondu, Thierry and Cloostermans, Ben and Dirkx, Yarince A. L. and Auman, Dorien and Cox, Stefaan and Verhoeven, Veronique and Hubens, Guy and Steenackers, Gunther and Tjalma, Wiebren A. A.},
title        = {Dynamic {InfraRed} Thermography ({DIRT}) in {DIEP}-flap breast reconstruction: A review of the literature},
journal      = {European Journal of Obstetrics \& Gynecology and Reproductive Biology},
year         = {2019},
volume       = {242},
pages        = {47--55},
doi          = {10.1016/j.ejogrb.2019.08.008}
}

@article{Wade2018,
author       = {Wade, Ryckie G. and Watford, James and Wormald, Justin C. R. and Bramhall, Russell J. and Figus, Andrea},
title        = {Perforator mapping reduces the operative time of {DIEP} flap breast reconstruction: A systematic review and meta-analysis of preoperative ultrasound, computed tomography and magnetic resonance angiography},
journal      = {Journal of Plastic, Reconstructive \& Aesthetic Surgery},
year         = {2018},
volume       = {71},
number       = {4},
pages        = {468--477},
doi          = {10.1016/j.bjps.2017.12.012},
pmid         = {29289500}
}

@inproceedings{Moderhak2016,
author       = {Moderhak, M. and Ko{\l}acz, Sz. and Jankau, Jerzy and Juchniewicz, T.},
title        = {An Active Dynamic Thermography ({ADT}) method for {TRAM} flap blood perfusion mapping in breast reconstruction},
booktitle    = {Proceedings of the 13th Quantitative InfraRed Thermography Conference (QIRT 2016)},
year         = {2016},
pages        = {580--581},
doi          = {10.21611/qirt.2016.089}
}

@article{DeWeerd2009,
author       = {de Weerd, Louis and Weum, Sven and Mercer, James B.},
title        = {The Value of Dynamic Infrared Thermography ({DIRT}) in Perforator Selection and Planning of Free {DIEP} Flaps},
journal      = {Annals of Plastic Surgery},
year         = {2009},
volume       = {63},
number       = {3},
pages        = {274--279},
doi          = {10.1097/SAP.0b013e318190321e},
pmid         = {19700958}
}

@article{Weum2016,
author       = {Weum, Sven and Mercer, James B. and de Weerd, Louis},
title        = {Evaluation of dynamic infrared thermography as an alternative to {CT} angiography for perforator mapping in breast reconstruction: A clinical study},
journal      = {BMC Medical Imaging},
year         = {2016},
volume       = {16},
number       = {1},
pages        = {43},
doi          = {10.1186/s12880-016-0144-x},
pmid         = {27421763},
pmcid        = {PMC4947300}
}

@article{Tenorio2011,
author       = {Tenorio, Xavier and Mahajan, Ajay L. and Elias, Badwi and Schaepkens van Riempst, Julie and Wettstein, Reto and Harder, Yves and Pittet, Brigitte},
title        = {Locating Perforator Vessels by Dynamic Infrared Imaging and Flow Doppler With No Thermal Cold Challenge},
journal      = {Annals of Plastic Surgery},
year         = {2011},
volume       = {67},
number       = {2},
pages        = {143--146},
doi          = {10.1097/SAP.0b013e3181ef6da3},
pmid         = {21407070}
}

@article{Meier2024,
author       = {Meier, Eva L. and Ulrich, Dietmar J. O. and Hummelink, Stefan},
title        = {Projected augmented reality in {DIEP} flap breast reconstruction: Projecting perforators on the skin using dynamic infrared thermography},
journal      = {Journal of Plastic, Reconstructive \& Aesthetic Surgery},
year         = {2024},
volume       = {94},
pages        = {83--90},
doi          = {10.1016/j.bjps.2024.05.001}
}

@article{Clarys2026_B,
  author       = {Clarys, Warre and Hotome, Valentine and Hummelink, Stefan and Verspeek, Simon and Verhoeven, Veronique and Tjalma, Wiebren A. A. and Steenackers, Gunther and Thiessen, Filip},
  title        = {Dynamic infrared thermography in free flap surgery: A systematic review},
  journal      = {JPRAS Open},
  year         = {2026},
  volume       = {48},
  pages        = {281--304},
  doi          = {10.1016/j.jpra.2025.11.027},
  note         = {Published online 27 November 2025}
}

@article{GagnonBlondeel2006,
  author       = {Gagnon, A. R. and Blondeel, P. N.},
  title        = {Colgajos de perforantes de las arterias epig{\'a}stricas inferiores profunda y superficial},
  journal      = {Cirug{\'i}a Pl{\'a}stica Ibero-Latinoamericana},
  year         = {2006},
  volume       = {32},
  number       = {4},
  pages        = {243--258},
  doi          = {10.4321/s0376-78922006000400004},
  language     = {Spanish},
  titleaddon   = {Deep and superficial inferior epigastric artery perforator flaps}
}

@article{Chawla2015,
  author       = {Chawla, Kunal and Gupta, Tulika and Tubbs, R. Shane and Aggarwal, Anjali and Sahni, Daisy and Chauhan, Randhir S. and Chawla, Rahul},
  title        = {A cadaveric study on abdominal wall perforators from the deep inferior epigastric artery: application to flap surgery},
  journal      = {European Journal of Anatomy},
  year         = {2015},
  volume       = {19},
  number       = {4},
  pages        = {315--322},
  url          = {https://www.eurjanat.com/data/pdf/eja.150118kc.pdf}
}

@article{Rupra2025_CTAmeta,
  author       = {Rupra, Roshan S. and Ruccia, Francesca and Daneshi, Kian and Aftab, Fatema and Yousif, Yousif F. and Khan, Gul R. and Dehnadi, Sina and AlSaidi, Yaqoob H. and Maggialetti, Nicola and Lorusso, Giovanni and Yan, Maria and Khajuria, Ankur},
  title        = {A systematic review and meta-analysis on computed tomography angiography mapping for deep inferior epigastric perforator flap breast reconstruction},
  journal      = {Frontiers in Oncology},
  year         = {2025},
  volume       = {15},
  pages        = {1600476},
  doi          = {10.3389/fonc.2025.1600476}
}

@article{Kim2025_CTAplanning,
  author       = {Kim, Jisu and Woo, Kyong-Je and Mun, Goo-Hyun},
  title        = {Making the most of computed tomography imaging in preoperative assessment and planning of abdomen-based flap breast reconstruction},
  journal      = {Plastic and Aesthetic Research},
  year         = {2025},
  volume       = {12},
  pages        = {17},
  doi          = {10.20517/2347-9264.2024.157}
}

@article{Bajus2023_CDUvsCTA,
  author       = {Bajus, Adam and Streit, Libor and Kubek, Tom{\'a}{\v{s}} and Nov{\'a}k, Adam and Van{\'i}{\v{c}}ek, Ji{\v{r}}{\'i} and {\v{S}}ediv{\'y}, Ond{\v{r}}ej and Berke{\v{s}}, Andrej and Can Bayezid, K. and Kunovsk{\'y}, Luk{\'a}{\v{s}} and Dra{\v{z}}an, Lubo{\v{s}}},
  title        = {Color Doppler ultrasound versus {CT} angiography for {DIEP} flap planning: A randomized controlled trial},
  journal      = {Journal of Plastic, Reconstructive \& Aesthetic Surgery},
  year         = {2023},
  volume       = {86},
  pages        = {48--57},
  doi          = {10.1016/j.bjps.2023.07.042}
}

@article{Lacey2025_XRreview,
  author       = {Lacey, Hester and Khan, Anas and Dheansa, Baljit},
  title        = {Extended reality for mapping perforator-based flaps in breast reconstruction: a systematic review and meta-analysis},
  journal      = {JPRAS Open},
  year         = {2025},
  volume       = {44},
  pages        = {269--283},
  doi          = {10.1016/j.jpra.2025.02.011}
}

@article{Kleiss2024_PerfusionMulti,
  author       = {Kleiss, Simone F. and Michi, Marlies and Schuurman, Simone N. and de Vries, Jean-Paul P. M. and Werker, Paul M. N. and de Jongh, Steven J.},
  title        = {Tissue perfusion in {DIEP} flaps using Indocyanine Green Fluorescence Angiography, Hyperspectral imaging, and Thermal imaging},
  journal      = {JPRAS Open},
  year         = {2024},
  volume       = {41},
  pages        = {61--74},
  doi          = {10.1016/j.jpra.2024.04.007}
}

@article{Alberti2025_ThermoDevice,
  author       = {Alberti, Andrea S. M. and Brucato, Davide and Parodi, Corrado and Schmauss, Daniel and Harder, Yves and Schweizer, Riccardo},
  title        = {Preoperative detection of perforating vessels in autologous abdominal flap-based breast reconstruction using a smartphone-based thermography device},
  journal      = {JPRAS Open},
  year         = {2025},
  volume       = {46},
  pages        = {461--472},
  doi          = {10.1016/j.jpra.2025.09.034}
}

@article{Sonda2022_Microsurgery,
  author       = {Sonda, Regina and Pandis, Laura and Bassetto, Franco and Marchica, Paolo and Messana, Francesco and Tiengo, Cesare and Andres, Anto Luigi and Brambullo, Tito and Vindigni, Vincenzo},
  title        = {Deep inferior epigastric perforator flap preoperative planning: A comparative analysis between dynamic infrared thermography, computerized tomography angiography, and hand-held Doppler},
  journal      = {Microsurgery},
  year         = {2022},
  volume       = {42},
  number       = {7},
  pages        = {649--658},
  doi          = {10.1002/micr.30927}
}

\end{document}